\documentclass[conference,letterpaper]{IEEEtran}

\usepackage[utf8]{inputenc} 
\usepackage[T1]{fontenc}
\usepackage{url}
\usepackage{ifthen}
\usepackage{cite}
\usepackage[cmex10]{amsmath} 

\usepackage{graphicx}
\usepackage{amsmath,amssymb,bm}
\usepackage{amsfonts}
\usepackage{mathtools}
\usepackage{tikz}

\newtheorem{definition}{Definition}
\newtheorem{lemma}{Lemma}
\newtheorem{theorem}{Theorem}

\newtheorem{cor}{Corollary}
\usepackage{caption}

\usepackage{booktabs, multirow} 
\usepackage{soul}
\usepackage{pgfplots}
\pgfplotsset{width=9cm,compat=1.9}

\begin{document}
\title{Polar Coded Computing:\\ The Role of the Scaling Exponent} 




\author{%
  \IEEEauthorblockN{Dorsa~Fathollahi\IEEEauthorrefmark{1},
                    and Marco~Mondelli\IEEEauthorrefmark{2}}
  \IEEEauthorblockA{\IEEEauthorrefmark{1}%
  Stanford University, Stanford, CA 94305, USA,
  dorsafth@stanford.edu}
  \IEEEauthorblockA{\IEEEauthorrefmark{2}%
Institute of Science and Technology (IST) Austria,
Klosterneuburg, Austria,
                     marco.mondelli@ist.ac.at}

}

\maketitle

\begin{abstract}
We consider the problem of coded distributed computing using polar codes. The average execution time of a coded computing system is related to the error probability for transmission over the binary erasure channel in recent work by Soleymani, Jamali and Mahdavifar, where the performance of binary linear codes is investigated. In this paper, we focus on polar codes and unveil a connection between the average execution time and the \emph{scaling exponent} $\mu$ of the family of codes. The scaling exponent has emerged as a central object in the finite-length characterization of polar codes, and it captures the speed of convergence to capacity. In particular, we show that \emph{(i)} the gap between the normalized average execution time of polar codes and that of optimal MDS codes is $O(n^{-1/\mu})$, and \emph{(ii)} this upper bound can be improved to roughly $O(n^{-1/2})$ by considering polar codes with large kernels. We conjecture that these bounds could be improved to $O(n^{-2/\mu})$ and $O(n^{-1})$, respectively, and provide a heuristic argument as well as numerical evidence supporting this view. 
\end{abstract}


\section{Introduction}
\label{sec:intro}

In recent years, there has been a high demand for data processing at large scale. This has motivated a paradigm shift towards distributed computing, where a large scale computation is spread over $n$ workers. A central issue is that some workers may finish later than others or crash, and a common solution is to add redundancy: coded distributed computing breaks a given large computational job into $k$ smaller tasks and adds $n-k$ redundant tasks, so that the computation is performed over $n$ workers. As a result, the final output is available even if we receive only part of the outputs. 

Applications of coding theory to distributed computing have been studied extensively, see e.g. \cite{stragglerproof-Baharav,lee2017speeding,polynomial2017salman,Mallick2019RatelessCF,Severinson2019BlockDiagonalAL,PolarCodedComputingMert,pilanci2021computational,soleymani2021coded} and references therein. Product codes are applied to distributed matrix multiplication in \cite{stragglerproof-Baharav}. In \cite{lee2017speeding}, maximum distance separable (MDS) codes are used to speed up matrix multiplication, and for matrix multiplication in a large finite field polynomial codes are employed in \cite{polynomial2017salman}. Luby Transform (LT) codes are proposed in \cite{Mallick2019RatelessCF,Severinson2019BlockDiagonalAL} for coded computation, 
and the application of polar codes to serverless computing is considered in 
\cite{PolarCodedComputingMert,pilanci2021computational}. In \cite{soleymani2021coded}, the average execution time of a coded computing system is related to the error probability of the linear code used to add redundancy. Although MDS codes are optimal in terms of average delay, they suffer from numerical stability and high decoding complexity, hence binary linear codes are studied in \cite{soleymani2021coded}: the authors characterize the performance gap between MDS and random binary linear codes and provide a numerical analysis of systems based on polar and Reed-Muller codes. 

In this paper, we analyze the performance of polar codes for coded distributed computing.  
Polar codes provably achieve capacity for any binary memoryless symmetric (BMS) channel with quasi-linear encoding and decoding complexity \cite{PolarcodeArikan}.
The code construction can be performed with linear complexity \cite{TV13con, RHTT} and, by exploiting a partial order between the synthetic channels, the construction complexity becomes sub-linear \cite{mondelli2018construction}. The error probability under successive cancellation (SC) decoding scales as $2^{-\sqrt{n}}$ \cite{ArT09}, $n$ being the block length, and polar codes are not affected by error floors \cite{unified-scale-MM}. Because of their attractive properties, polar codes have been recently adopted for the 5G wireless communications standard \cite{3gpp_polar}. 

A central object of interest in the analysis of polar codes has been the speed of convergence of the rate to capacity
\cite{FiniteLengthScaling-Kasra,improvedFiniteScaling-Goldin,speedpolarization-Guruswami,scalingExpList-Marco,unified-scale-MM,largeKernels-Marco,nearOptConv-Guruswami}. This line of work shows that the gap to capacity scales with the block length as $n^{-1/\mu}$. The parameter $\mu$ is called the \emph{scaling exponent}, it depends on the transmission channel and, for the special case of the binary erasure channel (BEC), $\mu \approx 3.63$ \cite{FiniteLengthScaling-Kasra, unified-scale-MM}. Furthermore, by using large polarization kernels, it is possible to approach the optimal scaling exponent $\mu=2$ \cite{largeKernels-Marco, nearOptConv-Guruswami}.

The goal of this paper is to characterize the average execution time of a coded computing system using polar codes via their scaling exponent. In particular, our main contributions can be summarized as follows:
\begin{itemize}
    \item In Section \ref{sec:avg}, we prove an upper bound on the average execution time of polar codes (Theorem \ref{thm:main}) and, as a consequence, we show that the normalized gap between polar codes and the optimal MDS codes is $O(n^{-1/\mu})$, $\mu$ being the scaling exponent (Corollary \ref{cor:gap}). 

    \item In Section \ref{sec:ker}, we show that the normalized gap can be improved to roughly $O(n^{-1/2})$ by considering polar codes with large kernels (Theorem \ref{thm:main_kernel}). 
    
    \item In Section \ref{sec:heu}, we consider codes of rate $R=1/2$ and provide a heuristic argument suggesting that, for polar codes with scaling exponent $\mu$, the normalized gap should scale as $n^{-2/\mu}$. The argument is also validated by numerical evidence. Hence, as polar codes with large kernels approach the optimal $\mu=2$ \cite{largeKernels-Marco, nearOptConv-Guruswami}, this would lead to a gap scaling as $n^{-1}$, which matches the performance of random binary linear codes. 
\end{itemize}



\section{Preliminaries}
\label{sec:def}


\subsection{Distributed Computing System Model}

We consider a distributed computing model with $n$ workers having the same computational capabilities, as in \cite{soleymani2021coded}. For the $i$-th worker, the run-time for performing a task is represented via a shifted exponential random variable $T_i$ \cite{lee2017speeding,reisizadeh2019coded,liang2014tofec}. Thus, dividing the computation equally among $k$ workers results in the following cumulative distribution function:
\begin{equation}
    \mathbb P(T_i\leq t ) = 1 - \exp (-\mu_{\rm s} (kt - 1 )), \quad\forall \, t\geq 1/k,
    \label{eq:shiftedexp}
\end{equation}
where $\mu_{\rm s}$ is the straggling parameter corresponding to each worker. Hence, the probability that the task is not completed until time $t\geq 1/k$ is given by 
\begin{equation}
    \epsilon(t) = \mathbb P(T_i \geq t) = \exp(-\mu_{\rm s}(kt-1)).
    \label{eq:erasureprob}
\end{equation}
We can now relate the problem of computing a task divided into $k$ parts over $n$ servers to the problem of transmitting $k$ symbols via an $(n, k)$ linear code over $n$ i.i.d. BECs with erasure probability $\epsilon(t)$. Let $T$ be the (random) execution time, defined as the time at which a decodable set of tasks is obtained from the workers. Then, the error probability of the $(n, k)$ code is related to the average execution time $\mathbb E[T]$ via the following lemma.



\begin{lemma}[Average execution time and error probability, Lemma 1 in \cite{soleymani2021coded}]  The average execution time $T_{\rm avg}$ of a coded distributed computing system using an $(n,k)$ linear code is given by 
\label{lemma:time-average-exec}
\begin{align}
    T_{\rm avg}\triangleq \mathbb{E}[T] = &  \int_{0}^{\infty}\hspace{-0.5em} P_e(\epsilon(\tau),n) {\rm d}\tau
    = \frac{1}{k} + \frac{1}{\mu_{\rm s} k} \int_0^1 \frac{P_e(\epsilon,n)}{\epsilon}{\rm d}\epsilon,
    \label{eq:Tavg}
\end{align}
where $\epsilon(\tau)$ is defined in \eqref{eq:erasureprob} and $P_e(\epsilon(\tau),n)$ denotes the error probability of the $(n, k)$ code for the transmission over a BEC$(\epsilon(\tau))$.
\end{lemma}


\subsection{Channel Polarization}

Channel polarization is based on mapping two identical copies of the transmission channel $W$ into a pair of ``synthetic'' channels $W^{+}$ and $W^{-}$, such that $W^+$ is strictly better and $W^-$ is strictly worse than $W$. Here, we will focus on the special case in which $W$ is a BEC, since this is the relevant setting for coded computing. When $W$ is a BEC$(\epsilon)$, we have that $W^+$ is a BEC$(\epsilon^2)$ and $W^-$ is a BEC$(2\epsilon-\epsilon^2)$. By repeating $m$ times this operation, we map $n\triangleq 2^m$ identical copies of $W$ into the synthetic channels $W_m^{(i)}$ ($i\in \{1, \ldots, n\}$). Consider a sequence of channels $W_m$ obtained by picking uniformly at random one of the synthetic channels $W_m^{(i)}$, and let $Z_m(W)$ be the random process that tracks the erasure probability of $W_m$. Then, 
\begin{equation}\label{eq:Zm}
Z_{m}= \begin{cases}Z_{m-1}^2, & \text {w.p. } 1 / 2, \\ 
2Z_{m-1}-Z_{m-1}^2, & \text {w.p. } 1 / 2,\end{cases}
\end{equation}
with initialization $Z_0=\epsilon$, $\epsilon$ being the erasure probability of the original channel $W$. The synthetic channels $W_m^{(i)}$ are all BECs and they polarize in the sense that, as $m$ grows, their erasure probabilities are either close to $0$ (\emph{noiseless} channels) or close to $1$ (\emph{noisy} channel). Formally, as $m
\to\infty$, $Z_m$ converges almost surely to a random variable $Z_\infty$ such that $\mathbb P(Z_\infty=0)=1-\epsilon$ and $\mathbb P(Z_\infty=1)=\epsilon$.

Now, the idea is to use the noiseless channels to transmit the information, while the noisy channels contain a fixed sequence shared between encoder and decoder (e.g., the all-zero sequence). In particular, in order to construct a polar code of block length $n$ and rate $R$ for transmission over $W=$ BEC($\epsilon$), we apply $m=\log_2 n$ steps of polarization to $W$ and select the $nR$ synthetic channels $W_m^{(i)}$ with the smallest erasure probabilities $Z_m^{(i)}$. We denote by $\mathcal I \subset \{1, \ldots, n\}$ the corresponding set of indices ($|\mathcal I|=nR$). Then, the error probability $P_e(\epsilon, n)$ of the polar code can be bounded as (see e.g. Section III of \cite{SATISH2010})
\begin{equation}\label{eq:ubPe}
  \max_{i\in \mathcal I} Z_m^{(i)} \le  P_e(\epsilon, n)\le \sum_{i\in\mathcal I}Z_m^{(i)}.
\end{equation}
In \cite{PolarcodeArikan}, it is shown that, as $n\to\infty$, the upper bound in \eqref{eq:ubPe} vanishes for any $R<I(W)=1-\epsilon$, thus implying that polar codes achieve capacity.

\subsection{Scaling Exponent}

The scaling exponent captures the speed of convergence of the rate to the channel capacity by quantifying how the gap to capacity scales as a function of the block length.

\begin{definition}[Upper bound on scaling exponent of BEC] We say that $\mu$ is an upper bound on the scaling exponent of the BEC, if there exists a function $h(x):[0,1] \rightarrow[0,1]$ such that $h(0)=h(1)=0$, $h(x)>0$ for any $x \in(0,1)$, and
$$
\sup _{x \in(0,1)} \frac{h(x^{2})+h(2 x-x^{2})}{2 h(x)}<2^{-1 / \mu} .
$$
\label{def:scaling-upper-bound}
\end{definition}

Using the definition above, in \cite{unified-scale-MM} it is proved that, if $\mu$ is an upper bound for the scaling exponent, then the gap to capacity $I(W) - R$ scales with the block length as $n^{-1/\mu}$. Furthermore, by constructing a suitable $h(x)$, in \cite{unified-scale-MM} it is also shown that $\mu =3.639$ is a valid upper bound for the BEC.

\subsection{Polar Codes with Large Kernels}

Conventional polar codes are obtained from the kernel $\begin{bmatrix}
1     & 0  \\
1     & 1
\end{bmatrix}$, in the sense that their generator matrix is given by the rows of $K^{\otimes m}$ (here, $K^{\otimes m}$ denotes the $m$-th Kronecker power of $K$) corresponding to the most reliable synthetic channels. In \cite{korada2010polar}, it was shown that capacity-achieving polar codes can be constructed from any kernel $K$ that is an $\ell\times \ell$ non-singular binary matrix, which cannot be transformed into an upper triangular matrix under any column permutations. In particular, given an $\ell\times \ell$ kernel $K$, we map $n\triangleq \ell^m$ identical copies of the transmission channel $W$ into the synthetic channels $W_m^{(i)}$ ($i\in \{1, \ldots, n\}$). If $W$ is a BEC, then the channels $\{W_m^{(i)}\}_{i=1}^n$ are all BECs. Their erasure probabilities are denoted by $\{Z_m^{(i)}\}_{i=1}^n$ and they are tracked by the random process $Z_m(W)$, which is defined as
\begin{equation}\label{eq:Bmkernels}
    Z_{m}=f_{K, B_{m-1}}(Z_{m-1}),
\end{equation}
where $B_{m-1}$ is drawn uniformly in $\{1, \ldots, \ell\}$ and the set of polynomials $\{f_{K, i}\}_{i=1}^\ell$ characterizes the polarization behavior of the kernel $K$ (for details and a formal definition, see e.g. (57) in \cite{largeKernels-Marco}). 
The error probability $P_e(\epsilon, n)$ of the resulting polar code can again be bounded as in \eqref{eq:ubPe}. By carefully analyzing the behavior of the polynomials $\{f_{K, i}\}_{i=1}^\ell$, when $K$ is a random non-singular binary matrix, in \cite{largeKernels-Marco} it is shown that the gap to capacity  of the polar code obtained from $K$ approaches the optimal scaling $n^{-1/2}$ as $\ell\to\infty$. These results were then generalized to the transmission over any BMS channel in \cite{nearOptConv-Guruswami}.

\section{Average Execution Time for Polar Codes}\label{sec:avg}


Our characterization of the average execution time via the scaling exponent exploits an intermediate result from \cite{unified-scale-MM}: we count the fraction of synthetic channels $W_m^{(i)}$ whose erasure probability is polynomially small in $n=2^m$, and we show that this fraction is at least the channel capacity $1-\epsilon$ \emph{minus} a term scaling as $n^{-1/\mu}=2^{-m/\mu}$, where $\mu$ is the scaling exponent. 

\begin{lemma}[Bound on $Z_m$]
Let $\mu\ge 2$ be an upper bound on the scaling exponent of the BEC in the sense of Definition \ref{def:scaling-upper-bound}, and let $Z_m$ be the random process tracking the erasure probability of $W_m$, where $W_0$ is a BEC$(\epsilon)$. Then, 
\begin{align}
\mathbb{P}&\left(Z_{m} \leq 2^{-10m}\right) \geq 1-\epsilon-c_1 2^{-m/\mu},\label{eq:z1n}
\end{align}
where $c_1$ is a numerical constant independent of $m, \epsilon$.
\label{lemma:polynomial-decay}
\end{lemma}
The proof of Lemma \ref{lemma:polynomial-decay} follows by setting $\nu=9$ into (35) of \cite{unified-scale-MM}, where $\rho$ is given by (33) and $\alpha$ is a factor $10$ smaller than in (31).
At this point, we are ready to state and prove our main result on the average execution time.

\begin{theorem}[$T_{\rm avg}$ for polar codes]\label{thm:main}
Let $\mu\ge 2$ be an upper bound on the scaling exponent of the BEC in the sense of Definition \ref{def:scaling-upper-bound}. Fix $R\in (0, 1)$ and a sufficiently large $n$, and define
\begin{equation}\label{eq:epsstar}
    \epsilon^* = 1-R-c_1\,n^{-1/\mu},
\end{equation}
where $c_1$ is the constant in \eqref{eq:z1n}. Let $T_{\rm avg}^{\rm polar}(n, R)$ be the average execution time of a coded distributed computing system using a polar code of block length $n$ and rate $R=k/n$, which is constructed for transmission over a BEC$(\epsilon^*)$. Then,
\begin{equation}\label{eq:claimthm}
    T_{\rm avg}^{\rm polar}(n, R) \leq \frac{1}{n\,R} + \frac{1}{\mu_{\rm s} n\,R} \left(-\ln\left(1-R\right)+c \,n^{-1/\mu}\right),
\end{equation}
where $c$ is a numerical constant independent of $n$.
\end{theorem}

\begin{IEEEproof}[Proof of Theorem \ref{thm:main}]
An application of Lemma \ref{lemma:time-average-exec} gives that
\begin{equation}\label{eq:Tavgint}
    T_{\rm avg}^{\rm polar}(n, R)    = \frac{1}{n\, R} + \frac{1}{\mu_{\rm s} n\, R} \int_0^1 \frac{P_e(\epsilon,n)}{\epsilon}{\rm d}\epsilon.
\end{equation}
To upper bound $\int_0^1 \frac{P_e(\epsilon,n)}{\epsilon}{\rm d}\epsilon$, we divide the integration domain into three intervals and bound each piece individually. 

First, pick $\epsilon \in [0,1/n^3]$ and note that $P_e(\epsilon,n)$ is trivially upper bounded by the probability that there is at least $1$ erasure, which is equal to $g(\epsilon)\triangleq 1-(1-\epsilon)^n$. Note that $g(0)=0$ and, for all $\epsilon\in [0, 1]$, $g'(\epsilon)=n(1-\epsilon)^{n-1}\le n$. Thus, $g(\epsilon)\le n\epsilon$ and, consequently, $P_e(\epsilon,n)\le n \epsilon$. Hence,
\begin{equation}\label{eq:int1}
  \int_{0}^{\frac{1}{n^3}} \frac{P_e(\epsilon,n)}{\epsilon}{\rm d}\epsilon \leq \frac{1}{n^2}.
\end{equation}

Next, let $\epsilon \in [1/n^3 , \epsilon^*]$, where $\epsilon^*$ is given by \eqref{eq:epsstar}. As the family of BECs is ordered by degradation \cite{richardson2008modern}, we have that 
\begin{equation}\label{eq:deg1}
 P_e(\epsilon, n)\le P_e(\epsilon^*, n).   
\end{equation}
Recall that the polar code is obtained by polarizing a BEC($\epsilon^*$), and let $\mathcal I \subset \{1, \ldots, n\}$ denote the corresponding set of $nR$ indices containing the information bits. By applying Lemma \ref{lemma:polynomial-decay_kernel} with initial condition $W_0=$ BEC($\epsilon^*$), we have that the fraction of synthetic channels whose erasure probability is at most $2^{-10m}= n^{-10}$ is lower bounded by
\begin{equation}
    1-\epsilon^*-c_1 2^{-m/\mu}=R,
\end{equation}
where we have used the definition \eqref{eq:epsstar} of $\epsilon^*$. Therefore, from the upper bound in \eqref{eq:ubPe}, we obtain that
\begin{equation}\label{eq:deg2}
    P_e(\epsilon^*, n)\le n^{-9},
\end{equation}
where we have also used that $|\mathcal I|=nR\le n$. Hence,
\begin{equation}\label{eq:int2}
\begin{aligned}
  \int_{\frac{1}{n^3}}^{\epsilon^*} \frac{P_e(\epsilon,n)}{\epsilon}{\rm d}\epsilon &\leq   \int_{\frac{1}{n^3}}^{\epsilon^*} \frac{1}{n^9 \epsilon} {\rm d}\epsilon \le\int_{\frac{1}{n^3}}^{\epsilon^*} \frac{1}{n^6} {\rm d}\epsilon\le \frac{1}{n^6}.
\end{aligned}
\end{equation}
Here, in the first inequality, we combine \eqref{eq:deg1} and \eqref{eq:deg2}; and in the second inequality, we use that $\epsilon\ge 1/n^3$.

Finally, let $\epsilon\in [\epsilon^*, 1]$. In this interval, we use the trivial upper bound $P_e(\epsilon, n)\le 1$. Hence,
\begin{equation}\label{eq:int3}
\begin{aligned}
   \int_{\epsilon^*}^1 \frac{P_e(\epsilon,n)}{\epsilon}{\rm d}\epsilon & \leq   \int_{\epsilon^*}^1 \frac{1}{\epsilon} {\rm d}\epsilon = -\ln(\epsilon^*).
\end{aligned}
\end{equation}
By combining \eqref{eq:epsstar}, \eqref{eq:int1}, \eqref{eq:int2} and \eqref{eq:int3}, we have that
\begin{equation}\label{eq:combint}
\begin{aligned}
     \int_0^1 \frac{P_e(\epsilon,n)}{\epsilon}{\rm d}\epsilon&\leq \frac{1}{n^2}+\frac{1}{n^6}- \ln(1-R-c_1n^{-1/\mu})   \\
     &\le \frac{1}{n^2}+\frac{1}{n^6}- \ln(1-R)+\frac{2c_1}{1-R}n^{-1/\mu}\\ 
     &\le - \ln(1-R)+c n^{-1/\mu}. 
\end{aligned}
\end{equation}
Here, the second inequality holds as $-\ln(1-x)\le 2x$ for $x\in [0, 1/2]$ and $\frac{c_1}{1-R}n^{-1/\mu}\in [0, 1/2]$ for sufficiently large $n$; and, to obtain the third inequality, we use that $\mu\ge 2$ and pick $c=\frac{2c_1}{1-R}+2$. By combining \eqref{eq:Tavgint} and \eqref{eq:combint}, the desired result readily follows.
\end{IEEEproof}

Armed with this result, we can bound the gap between polar codes and MDS codes, which achieve the minimum average execution time (cf. Corollary 6 in \cite{soleymani2021coded}).

\begin{cor}[Gap to optimum for polar codes]\label{cor:gap}
Fix $R\in (0, 1)$ and a sufficiently large $n$. Let $T_{\rm avg}^{\rm MDS}(n, R)$ be the average execution time of a coded distributed computing system using an MDS code of block length $n$ and rate $R=k/n$. Let $T_{\rm avg}^{\rm polar}(n, R)$ be defined as in Theorem \ref{thm:main}. Then, 
\begin{equation}
    nT_{\rm avg}^{\rm polar}(n, R)-nT_{\rm avg}^{\rm MDS}(n, R) \leq c' n^{-1/\mu},
\end{equation}
where $c'$ is a numerical constant independent of $n$.
\end{cor}

\begin{IEEEproof}
Recall from (15) in \cite{soleymani2021coded} that
\begin{equation}\label{eq:TavgMDS}
    T_{\rm avg}^{\rm MDS}(n, R)=\frac{1}{n\,R}+\frac{1}{\mu_{\rm s}n\,R}\sum_{i=n(1-R)+1}^n\frac{1}{i}.
\end{equation}
We lower bound the harmonic series on the RHS of \eqref{eq:TavgMDS} as 
\begin{equation}\label{eq:TavgMDS2}
\begin{split}
    &\sum_{i=n(1-R)+1}^n\frac{1}{i}\ge \int_{n(1-R)+1}^n \frac{1}{t}{\rm d}t = \ln(n)-\ln(n(1-R)+1) \\
    &\hspace{2em}= -\ln(1-R+1/n)\ge -\ln(1-R)-\frac{1}{n(1-R)},
\end{split}
\end{equation}
where first inequality uses that $f(t)=1/t$ is decreasing and the last inequality uses that $\log(1+t)\le t$ for all $t\ge 0$. By combining \eqref{eq:TavgMDS} and \eqref{eq:TavgMDS2}, we deduce that  
\begin{equation}\label{eq:TavgMDS3}
    nT_{\rm avg}^{\rm MDS}(n, R) \ge \frac{1}{R}-\frac{1}{\mu_{\rm s}R}\ln(1-R)-\frac{1}{\mu_{\rm s}R(1-R)}\frac{1}{n}.
\end{equation}
Recall that $\mu\ge 2$ and pick $c'=c(\mu_{\rm s}R)^{-1}+(\mu_{\rm s}R(1-R))^{-1}$, where $c$ is the numerical constant in \eqref{eq:claimthm}. Thus, the claim follows by combining \eqref{eq:TavgMDS3} with the result of Theorem \ref{thm:main}. 
\end{IEEEproof}

We remark that random binary linear codes achieve a smaller gap with respect to the optimal MDS codes. In fact, $nT_{\rm avg}^{\rm BRC}(n, R)-nT_{\rm avg}^{\rm MDS}(n, R)$ is $O\left(\frac{\log_2 n}{n}\right)$, where $T_{\rm avg}^{\rm BRC}(n, R)$ denotes the average execution time for a random binary linear code of block length $n$ and rate $R$ (cf. Corollary 9 in \cite{soleymani2021coded}). Thus, a natural question is whether it is possible to reduce the gap between polar and MDS codes. We answer this question affirmatively in the next section, where we consider polar codes obtained from large random kernels.

\section{Extension to Polar Codes with Large Kernels}\label{sec:ker}

First, we consider the fraction of synthetic channels whose erasure probability is polynomially small in $n=\ell^m$, and we show that it is lower bounded by  the channel capacity $1-\epsilon$ \emph{minus} a term scaling roughly as $\ell^{-m/2}=n^{-1/2}$. This result is stated below, and it is a (slight) improvement over Theorem 3 in \cite{largeKernels-Marco}. Its proof is deferred to the appendix. 

\begin{lemma}[Bounds on $Z_m$ for large kernels]
Let $K$ be a kernel selected uniformly at random from all $\ell\times \ell$ non-singular binary matrices, and let $Z_m$ be the random process defined in \eqref{eq:Bmkernels} with initial condition $Z_0=\epsilon$. Fix a small constant $\delta>0$. Then, there exists $\ell_0(\delta)$ such that, for any $\ell>\ell_0(\delta)$ and for all $m\ge 1$, the following holds with high probability over the choice of $K$:
\begin{align}
\mathbb{P}&\left(Z_{m} \leq \ell^{-10m}\right) \geq 1-\epsilon-c_2 \ell^{-m/(2+\delta)},\label{eq:z3n}
\end{align}
where $c_2$ is a numerical constant independent of $m, \ell,\delta, \epsilon$.
\label{lemma:polynomial-decay_kernel}
\end{lemma}

At this point, we are ready to state our results for polar codes based on large kernels.

\begin{theorem}[$T_{\rm avg}$ and gap to optimum for polar codes with large kernels]\label{thm:main_kernel}
Fix  $R\in (0, 1)$, a small constant $\delta>0$, a sufficiently large $\ell$ and a sufficiently large $n$. Define
\begin{equation}\label{eq:epsstarken}
    \epsilon^* = 1-R-c_2\,n^{-1/(2+\delta)},
\end{equation}
where $c_2$ is the constant in \eqref{eq:z3n}. Let $K$ be a kernel selected uniformly at random from all $\ell\times \ell$ non-singular binary matrices, let $\mathcal C_K(n, R, \delta)$ be the polar code of block length $n=\ell^m$ and rate $R$ obtained from the kernel $K$ and constructed for transmission over a BEC($\epsilon^*$), and let $T_{\rm avg}^{K}(n, R, \delta)$ be the average execution time of a coded distributed computing system using $\mathcal C_K(n, R, \delta)$. Then, the following holds with high probability over the choice of $K$:
\begin{equation}\label{eq:claimthm_ker}
    T_{\rm avg}^{K}(n, R, \delta) \leq \frac{1}{n\,R} + \frac{1}{\mu_{\rm s} n\,R} \left(-\ln\left(1-R\right)+\tilde{c} \,n^{-1/(2+\delta)}\right),
\end{equation}
where $\tilde{c}$ is a numerical constant independent of $n$. Furthermore, let $T_{\rm avg}^{\rm MDS}(n, R)$ be the average execution time of a coded distributed computing system using an MDS code of block length $n$ and rate $R$. Then,
\begin{equation}\label{eq:claimthm_ker2}
    nT_{\rm avg}^{K}(n, R, \delta)-nT_{\rm avg}^{\rm MDS}(n, R) \leq \tilde{c}' n^{-1/(2+\delta)},
\end{equation}
where $\tilde{c}'$ is a numerical constant independent of $n$.
\end{theorem}
The proof of \eqref{eq:claimthm_ker} is analogous to that of Theorem \ref{thm:main}, the only difference being that the application of Lemma \ref{lemma:polynomial-decay} is replaced by the application of Lemma \ref{lemma:polynomial-decay_kernel}. The proof of \eqref{eq:claimthm_ker2} follows from \eqref{eq:claimthm_ker} and from the lower bound on $nT_{\rm avg}^{\rm MDS}(n, R)$ in \eqref{eq:TavgMDS3}. 

The upper bound in \eqref{eq:claimthm_ker2} roughly scales as $n^{-1/2}$, hence the result of Theorem \ref{thm:main_kernel} still does not match the performance of random binary linear codes. We conjecture that this is not due to a sub-optimality of polar codes using large kernels, but rather to a sub-optimality of our bounds. In fact, in the following section, we will give a simple heuristic argument suggesting that, for a family of polar codes of rate $R=1/2$ and scaling exponent $\mu$, the gap to optimum $nT_{\rm avg}^{\rm polar}(n, R)-nT_{\rm avg}^{\rm MDS}(n, R)$ is $O(n^{-2/\mu})$ (as opposed to $O(n^{-1/\mu})$ via Corollary \ref{cor:gap}). Since polar codes with large kernels approach $\mu= 2$, this improved bound would match the gap to optimum of random binary linear codes.

\section{Improving the Bounds: A Heuristic Argument and Numerical Experiments}\label{sec:heu}

We focus on the case $R=1/2$. Our heuristic argument relies on the assumption that the error probability $P_e(\epsilon, n)$ behaves as the lower bound $\max_{i\in \mathcal I} Z_m^{(i)}$ in \eqref{eq:ubPe}. We also assume the existence of a scaling law s.t. 
\begin{equation}\label{eq:scalass}
    \lim_{n\to \infty \,:\, n^{1/\mu}(1/2-\epsilon)=z } P_e(\epsilon, n)=f(z).
\end{equation}
Here, $\mu$ represents the \emph{scaling exponent} and $f$ is typically referred to as the \emph{mother curve}. The idea is that, as the block length diverges, the gap to capacity $1/2-\epsilon$ scales as $n^{-1/\mu}$, and in this regime the error probability tends to the limit $f(n^{1/\mu}(1/2-\epsilon))\triangleq f(z)$. In \cite{SATISH2010}, the authors provide numerical and rigorous evidence that the lower bound in \eqref{eq:ubPe} satisfies such a scaling law. Furthermore, the scaling exponent for the BEC is estimated as $\mu\approx 3.6261$, a value which is close to the rigorous upper bound of $3.639$ coming from \cite{unified-scale-MM}.

We will also use that the derivative of the mother curve $f'(\cdot)$ is even. To justify this, recall that $Z_m(\epsilon)$ is given by the recursion \eqref{eq:Zm} with initial condition $Z_0=\epsilon$, $\{Z_m^{(i)}(\epsilon)\}_{i=1}^n$ are the erasure probabilities of the synthetic channels obtained by polarizing a BEC($\epsilon$), and $\mathcal I(\epsilon)$ is the set of indices corresponding to the synthetic channels with the smallest erasure probabilities $Z_m^{(i)}(\epsilon)$. Similarly, $Z_m(1-\epsilon)$ is given by \eqref{eq:Zm} with initial condition $Z_0=1-\epsilon$, $\{Z_m^{(i)}(1-\epsilon)\}_{i=1}^n$ denote the erasure probabilities obtained from the polarization of a BEC($1-\epsilon$), and $\mathcal I(1-\epsilon)$ is the corresponding set of indices. Then, one can easily verify that $Z_m(1-\epsilon)=1-Z_m(\epsilon)$. As $R=1/2$, this implies that $\max_{i\in \mathcal I(1-\epsilon)} Z_m^{(i)}(1-\epsilon)= 1-\min_{i\in \mathcal I^c(\epsilon)} Z_m^{(i)}(\epsilon)$, where $\mathcal I^c(\epsilon)$ is the complement set of $\mathcal I(\epsilon)$. Furthermore, as $m\to\infty$, $\min_{i\in \mathcal I^c(\epsilon)} Z_m^{(i)}(\epsilon)\approx \max_{i\in \mathcal I(\epsilon)} Z_m^{(i)}(\epsilon)$. Therefore, assuming that the lower bound in \eqref{eq:ubPe} is tight, $P_e(1-\epsilon, n)=1-P_e(\epsilon, n)$, which under the scaling assumption \eqref{eq:scalass}, implies that $f'(\cdot)$ is even. 

The goal of our heuristic argument is to show that 
\begin{equation}\label{eq:heuend}
    \int_{1/4}^{3/4}  \frac{P_e(\epsilon, n)}{\epsilon}{\rm d}\epsilon =\ln(3/4)-\ln(1/2)+O(n^{-2/\mu}). 
\end{equation}
Note that, by using \eqref{eq:int1} and \eqref{eq:int2}, the integral of $P_e(\epsilon, n)/\epsilon$ over the interval $[0, 1/4]$ is $O(n^{-2})$. Furthermore, by using the trivial upper bound $P_e(\epsilon, n)\le 1$, the same integral over the interval $[3/4, 1]$ is upper bounded by $-\ln(3/4)$. Hence, \eqref{eq:heuend} implies that 
\begin{equation}
 T_{\rm avg}^{\rm polar}(n, R) \leq \frac{1}{n\,R} + \frac{1}{\mu_{\rm s} n\,R} \left(-\ln\left(1-R\right)+O( n^{-2/\mu})\right),   
\end{equation}
which implies that the gap to optimum for polar codes scales as $n^{-2/\mu}$, as desired. 

We now show how to obtain \eqref{eq:heuend}. By performing the change of variables $z=n^{1/\mu}(1/2-\epsilon)$ and using the scaling law assumption \eqref{eq:scalass}, we have 
\begin{equation}\label{eq:h1}
\begin{split}
    \int_{1/4}^{3/4}  \frac{P_e(\epsilon, n)}{\epsilon}{\rm d}\epsilon &= n^{-1/\mu}\int_{-n^{1/\mu}/4}^{n^{1/\mu}/4} \frac{P_e(1/2-z n^{-1/\mu}, n)}{1/2-z n^{-1/\mu}}{\rm d}z\\
    &\approx n^{-1/\mu}\int_{-n^{1/\mu}/4}^{n^{1/\mu}/4} \frac{f(z)}{1/2-z n^{-1/\mu}}{\rm d}z.
    \end{split}
\end{equation}
Next, we integrate by parts the RHS of \eqref{eq:h1}, which gives
\begin{equation}\label{eq:h2}
\begin{split}
    - &f(z)\ln(1/2-z n^{-1/\mu})\bigg|_{-n^{1/\mu}/4}^{n^{1/\mu}/4}\\    
 &+\int_{-n^{1/\mu}/4}^{n^{1/\mu}/4} f'(z)\ln(1/2-z n^{-1/\mu}){\rm d}z:= T_1+T_2.
\end{split}
\end{equation}
The term $T_1$ in the first line of \eqref{eq:h2} is upper bounded by $\ln(3/4)+O(n^{-9})$. In fact, $f(-n^{1/\mu}/4)\approx P_e(3/4, n)\approx 1$ and $f(n^{1/\mu}/4)\approx P_e(1/4, n)\le n^{-9}$, where the last inequality follows from \eqref{eq:deg1} and \eqref{eq:deg2}. Finally, by performing a Taylor expansion of $\ln(1/2-z n^{-1/\mu})$ around $1/2$, the term $T_2$ in the second line of \eqref{eq:h2} is upper bounded by 
\begin{equation}
\begin{split}
        \ln(1/2)\int_{-n^{1/\mu}/4}^{n^{1/\mu}/4} \hspace{-1.5em}f'(z){\rm d}z+2n^{-1/\mu}\int_{-n^{1/\mu}/4}^{n^{1/\mu}/4} \hspace{-1.5em}zf'(z){\rm d}z &+ O(n^{-2/\mu})\\
        =-\ln(1/2)&+O(n^{-2/\mu}),
\end{split}
\end{equation}
where we have used that the second term in the first line is $0$ (being the integral of an odd function over a symmetric interval). By summing up these bounds on $T_1$ and $T_2$, we obtain that the quantity in \eqref{eq:h2} is upper bounded by $\ln(3/4)-\ln(1/2)+O(n^{-2/\mu})$, which combined with \eqref{eq:h1} implies the desired result \eqref{eq:heuend}.

\begin{figure}
    \centering

\begin{tikzpicture}

\begin{axis}[
scale=1,
xmin=8,
xmax=24,
ymin=-16,
ymax=-5,
xmajorgrids=true,
grid style=dashed,
width=.45\textwidth,
xlabel={$\log_2(n)$},
ylabel={$\log_2 \big(g(n)-\ln(1-R) \big)$},
ylabel shift=-7,
legend cell align={left},
legend pos=north west,
set layers=standard,
legend style={
	column sep= 1mm,
	font=\fontsize{9pt}{9}\selectfont,
	draw=none,
	fill opacity=0.75,
	text opacity = 1,
},
clip mode=individual,
]

\foreach \i in {8,...,24}
{
\addplot[
color=gray,
dashed,
domain=18:24,
samples=10,
smooth,
forget plot,
]
{-0.5499407091606934 *x -1.9716412175367477};
}

\draw [color=gray, dashed, opacity=1] (60 , 70.03191623282166) -- (100 , 48.03428786639391) node[midway,above, sloped] {slope $-0.550536$};

\addplot[
color=blue,
mark=*,
thick
]
table {
8 -6.8159507312132295
9 -7.16617953356784
10 -7.666422653720387
11 -8.259250872153794
12 -8.690732371236399
13 -9.219590559464033
14 -9.70697656040705
15 -10.26479172340695
16 -10.790636410450015
17 -11.32363087566303
18 -11.871753939551517
19 -12.418491378731384
20 -12.972598596936209
21 -13.518669826389056
22 -14.070449967822032
23 -14.61969411811608
24 -15.17111494183291
};

\end{axis}
\end{tikzpicture}
\caption{Evaluation of the function $g(n)$ defined in \eqref{eq:gn} for polar codes of rate $R=1/2$ and block lengths from $2^8$ up to $2^{24}$. 
  }
\label{plt:lb}
\end{figure}
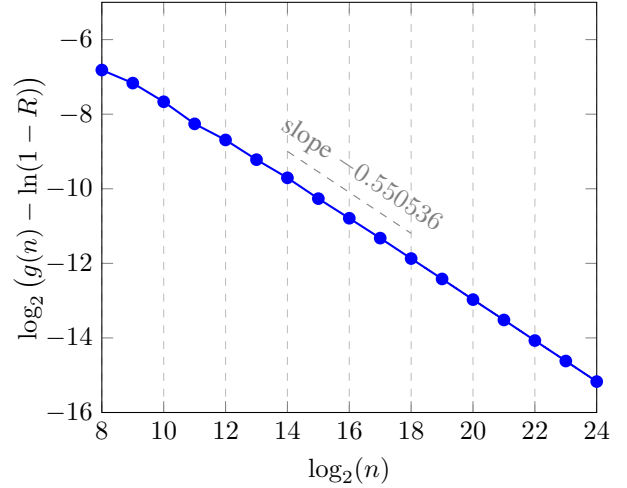

The scaling suggested by the heuristic argument above is also supported by numerical results. 
Let $g(n)$ be defined as 
\begin{equation}\label{eq:gn}
    g(n) \triangleq \int_0^1  \frac{\max_{i\in \mathcal I(\epsilon)} Z_m^{(i)}(\epsilon)}{\epsilon}d\epsilon\le \int_0^1 \frac{P_e(\epsilon,n)}{\epsilon}{\rm d}\epsilon,
\end{equation}
where the upper bound follows from \eqref{eq:ubPe}. The plot of Figure \ref{plt:lb} shows that $g(n)-\ln(1-R)$ scales as $n^{-2/\mu}$, where the numerical value of $\mu$ matches the scaling exponent of the BEC. 


\section{Conclusion}

In this paper, we consider the average execution time of a coded computing system using polar codes. We show that the performance gap between polar codes and the optimal MDS codes is upper bounded as $O(n^{-1/\mu})$, $n$ being the block length. Here, $\mu$ is the scaling exponent of polar codes for transmission over the BEC, and tight bounds have been proved on this quantity ($\mu \approx 3.63$). Next, we consider polar codes based on large kernels, which are known to approach the optimal scaling exponent $\mu=2$. For this family of codes, we prove that the gap to MDS codes scales roughly as $O(n^{-1/2})$. Finally, we conjecture that our bounds could be improved to $O(n^{-2/\mu})$ and $O(n^{-1})$, respectively, which would imply that polar codes with large kernels match the performance of binary random linear codes. Our conjecture is supported by a heuristic argument, as well as by numerical evidence. 


\section*{Acknowledgments}

D. Fathollahi and M. Mondelli were partially supported by the 2019 Lopez-Loreta Prize. The authors thank Hamed Hassani and Hessam Mahdavifar for helpful discussions. 

\newpage

\bibliographystyle{IEEEtran}
\bibliography{references}

\begin{thebibliography}{10}
\providecommand{\url}[1]{#1}
\csname url@samestyle\endcsname
\providecommand{\newblock}{\relax}
\providecommand{\bibinfo}[2]{#2}
\providecommand{\BIBentrySTDinterwordspacing}{\spaceskip=0pt\relax}
\providecommand{\BIBentryALTinterwordstretchfactor}{4}
\providecommand{\BIBentryALTinterwordspacing}{\spaceskip=\fontdimen2\font plus
\BIBentryALTinterwordstretchfactor\fontdimen3\font minus
  \fontdimen4\font\relax}
\providecommand{\BIBforeignlanguage}[2]{{%
\expandafter\ifx\csname l@#1\endcsname\relax
\typeout{** WARNING: IEEEtran.bst: No hyphenation pattern has been}%
\typeout{** loaded for the language `#1'. Using the pattern for}%
\typeout{** the default language instead.}%
\else
\language=\csname l@#1\endcsname
\fi
#2}}
\providecommand{\BIBdecl}{\relax}
\BIBdecl

\bibitem{stragglerproof-Baharav}
T.~Baharav, K.~Lee, O.~Ocal, and K.~Ramchandran, ``Straggler-proofing
  massive-scale distributed matrix multiplication with d-dimensional product
  codes,'' in \emph{IEEE International Symposium on Information Theory (ISIT)},
  2018, pp. 1993--1997.

\bibitem{lee2017speeding}
K.~Lee, M.~Lam, R.~Pedarsani, D.~Papailiopoulos, and K.~Ramchandran, ``Speeding
  up distributed machine learning using codes,'' \emph{IEEE Transactions on
  Information Theory}, vol.~64, no.~3, pp. 1514--1529, Mar. 2017.

\bibitem{polynomial2017salman}
Q.~Yu, M.~A. Maddah-Ali, and A.~S. Avestimehr, ``Polynomial codes: An optimal
  design for high-dimensional coded matrix multiplication,'' in \emph{Neural
  Information Processing Systems (NeurIPS)}, 2017, p. 4406–4416.

\bibitem{Mallick2019RatelessCF}
A.~Mallick, M.~Chaudhari, and G.~Joshi, ``Rateless codes for near-perfect load
  balancing in distributed matrix-vector multiplication,'' \emph{Proceedings of
  the ACM on Measurement and Analysis of Computing Systems}, vol.~3, pp. 1 --
  40, 2019.

\bibitem{Severinson2019BlockDiagonalAL}
A.~Severinson, A.~G. i~Amat, and E.~Rosnes, ``Block-diagonal and {LT} codes for
  distributed computing with straggling servers,'' \emph{IEEE Transactions on
  Communications}, vol.~67, no.~3, pp. 1739--1753, Mar. 2019.

\bibitem{PolarCodedComputingMert}
B.~Bartan and M.~Pilanci, ``Straggler resilient serverless computing based on
  polar codes,'' in \emph{57th Annual Allerton Conference on Communication,
  Control, and Computing (Allerton)}, 2019, pp. 276--283.

\bibitem{pilanci2021computational}
M.~Pilanci, ``Computational polarization: An information-theoretic method for
  resilient computing,'' \emph{IEEE Transactions on Information Theory}, to
  appear.

\bibitem{soleymani2021coded}
M.~Soleymani, M.~V. Jamali, and H.~Mahdavifar, ``Coded computing via binary
  linear codes: Designs and performance limits,'' \emph{IEEE Journal on
  Selected Areas in Information Theory}, vol.~2, no.~3, pp. 87--892, Sept.
  2021.

\bibitem{PolarcodeArikan}
E.~Arikan, ``Channel polarization: A method for constructing capacity-achieving
  codes for symmetric binary-input memoryless channels,'' \emph{IEEE
  Transactions on information Theory}, vol.~55, no.~7, pp. 3051--3073, 2009.

\bibitem{TV13con}
I.~Tal and A.~Vardy, ``How to construct polar codes,'' \emph{IEEE Transactions
  on Information Theory}, vol.~59, no.~10, pp. 6562--6582, Oct. 2013.

\bibitem{RHTT}
R.~Pedarsani, H.~Hassani, I.~Tal, and E.~Telatar, ``On the construction of
  polar codes,'' in \emph{IEEE International Symposium on Information Theory
  (ISIT)}, St. Petersberg, Russia, Aug. 2011, pp. 11--15.

\bibitem{mondelli2018construction}
M.~Mondelli, S.~H. Hassani, and R.~Urbanke, ``Construction of polar codes with
  sublinear complexity,'' \emph{IEEE Transactions on Information Theory},
  vol.~65, no.~5, pp. 2782--2791, May 2019.

\bibitem{ArT09}
E.~{Ar\i kan} and I.~E. {Telatar}, ``{On the rate of channel polarization},''
  in \emph{IEEE International Symposium on Information Theory (ISIT)}, Seoul,
  South Korea, July 2009, pp. 1493--1495.

\bibitem{unified-scale-MM}
M.~Mondelli, S.~H. Hassani, and R.~L. Urbanke, ``Unified scaling of polar
  codes: Error exponent, scaling exponent, moderate deviations, and error
  floors,'' \emph{IEEE Transactions on Information Theory}, vol.~62, no.~12,
  pp. 6698--6712, 2016.

\bibitem{3gpp_polar}
``Final report of {3GPP TSG RAN WG1} \#87 v1.0.0,'' {R}eno, USA, Nov. 2016.

\bibitem{FiniteLengthScaling-Kasra}
S.~H. Hassani, K.~Alishahi, and R.~L. Urbanke, ``Finite-length scaling for
  polar codes,'' \emph{IEEE Transactions on Information Theory}, vol.~60,
  no.~10, pp. 5875--5898, Oct. 2014.

\bibitem{improvedFiniteScaling-Goldin}
D.~Goldin and D.~Burshtein, ``Improved bounds on the finite length scaling of
  polar codes,'' \emph{IEEE Transactions on Information Theory}, vol.~60,
  no.~11, pp. 6966--6978, Nov. 2014.

\bibitem{speedpolarization-Guruswami}
V.~Guruswami and P.~Xia, ``Polar codes: {S}peed of polarization and polynomial
  gap to capacity,'' \emph{IEEE Transactions on Information Theory}, vol.~61,
  no.~1, pp. 3--16, Jan. 2015.

\bibitem{scalingExpList-Marco}
M.~Mondelli, S.~H. Hassani, and R.~Urbanke, ``Scaling exponent of list decoders
  with applications to polar codes,'' \emph{IEEE Transactions on Information
  Theory}, vol.~61, no.~9, pp. 4838--4851, Sept. 2015.

\bibitem{largeKernels-Marco}
A.~Fazeli, H.~Hassani, M.~Mondelli, and A.~Vardy, ``Binary linear codes with
  optimal scaling: Polar codes with large kernels,'' \emph{IEEE Transactions on
  Information Theory}, vol.~67, no.~9, pp. 5693--5710, Sept. 2021.

\bibitem{nearOptConv-Guruswami}
V.~Guruswami, A.~Riazanov, and M.~Ye, ``Arikan meets shannon: Polar codes with
  near-optimal convergence to channel capacity,'' in \emph{52nd Annual ACM
  SIGACT Symposium on Theory of Computing (STOC)}, 2020, p. 552–564.

\bibitem{reisizadeh2019coded}
A.~Reisizadeh, S.~Prakash, R.~Pedarsani, and A.~S. Avestimehr, ``Coded
  computation over heterogeneous clusters,'' \emph{IEEE Transactions on
  Information Theory}, vol.~65, no.~7, pp. 4227--4242, July 2019.

\bibitem{liang2014tofec}
G.~Liang and U.~C. Kozat, ``{TOFEC}: Achieving optimal throughput-delay
  trade-off of cloud storage using erasure codes,'' in \emph{IEEE INFOCOM
  2014-IEEE Conference on Computer Communications}, 2014, pp. 826--834.

\bibitem{SATISH2010}
S.~B. Korada, A.~Montanari, E.~Telatar, and R.~Urbanke, ``An empirical scaling
  law for polar codes,'' in \emph{2010 IEEE International Symposium on
  Information Theory}, 2010, pp. 884--888.

\bibitem{korada2010polar}
S.~B. Korada, E.~{\c{S}}a{\c{s}}o{\u{g}}lu, and R.~Urbanke, ``Polar codes:
  Characterization of exponent, bounds, and constructions,'' \emph{IEEE
  Transactions on Information Theory}, vol.~56, no.~12, pp. 6253--6264, Dec.
  2010.

\bibitem{richardson2008modern}
T.~Richardson and R.~Urbanke, \emph{Modern coding theory}.\hskip 1em plus 0.5em
  minus 0.4em\relax Cambridge university press, 2008.

\end{thebibliography}

\appendix

\begin{IEEEproof}[Proof of Lemma \ref{lemma:polynomial-decay_kernel}]
Let $g_\ell(z)=z^{1/\ln\ell}(1-z)^{1/\ln\ell}$. By combining Lemma 5 and Lemma 6 in \cite{largeKernels-Marco}, we have that, for any $m\ge 0$, with high probability over the choice of $K$,
\begin{equation}\label{eq:bdker}
    \mathbb E[g_\ell(Z_m)]\le \left(c_3\ell^{-1/2}\ln \ell\right)^m g_\ell(\epsilon),
\end{equation}
where $c_3$ is a numerical constant.
Then, the following chain of inequalities holds:
\begin{equation}\label{eq:bdmiddle}
\begin{split}
    \mathbb P(Z_m\in &[\ell^{-10m}, 1-\ell^{-10m}]) = \mathbb P(g_\ell(Z_m)\ge g_\ell(\ell^{-10m}))\\
    &\le \frac{\mathbb E[g_\ell(Z_m)]}{g_\ell(\ell^{-10m})}\\
    &\le \frac{\left(c_3\ell^{-1/2}\ln \ell\right)^m g_\ell(\epsilon)}{g_\ell(\ell^{-10m})}\\
    &\le \left(c_3\ell^{-1/2+10/\ln \ell}\ln \ell\right)^m\frac{1}{(1-\ell^{-10m})^{1/\ln \ell}} \\
    &\le 2\left(c_3\ell^{-1/2+10/\ln \ell}\ln \ell\right)^m.
\end{split}
\end{equation}
Here, in the first line we use the concavity of $g_\ell(\cdot)$ together with its symmetry around $1/2$; in the second line, we use Markov's inequality; in the third line, we use \eqref{eq:bdker}; in the fourth line, we use that $g_\ell(\epsilon)\le 1$ and that $g_\ell(\ell^{-10m})=\ell^{-10m/\ln \ell}(1-\ell^{-10m})^{1/\ln \ell}$; and in the fifth line, we use that $(1-\ell^{-10m})^{1/\ln \ell}\ge 1/2$ for $m\ge 1$ and $\ell\ge 3$. 

The rest of the argument follows the proof of Lemma 4 in \cite{largeKernels-Marco}, and it is briefly outlined below. Let us define 
\begin{equation}\label{eq:ABC}
\begin{split}
    A &\triangleq\mathbb P(Z_m\in [0, \ell^{-10m})),\\
    B &\triangleq\mathbb P(Z_m\in [ \ell^{-10m}, 1-\ell^{-10m}]),\\
    C &\triangleq\mathbb P(Z_m\in ( 1-\ell^{-10m}, 1)).\\
\end{split}
\end{equation}
Furthermore, let $A'$, $B'$, and $C'$ be the fraction of indices in $A$, $B$, and $C$, respectively, that will have a vanishing erasure probability, i.e.,
\begin{equation}\label{eq:ABCp}
\begin{split}
    A' &\triangleq\lim_{m'\to\infty}\mathbb P\left(Z_m\in (0, \ell^{-10m}), Z_{m+m'}\le \ell^{-m'}\right),\\
    B' &\triangleq\lim_{m'\to\infty}\mathbb P\left(Z_m\in [ \ell^{-10m}, 1-\ell^{-10m}], Z_{m+m'}\le \ell^{-m'}\right),\\
    C' &\triangleq\lim_{m'\to\infty}\mathbb P\left(Z_m\in ( 1-\ell^{-10m}, 1), Z_{m+m'}\le \ell^{-m'}\right).\\
\end{split}
\end{equation}
The existence of the limits in \eqref{eq:ABCp} -- and therefore the well-posedness of $A'$, $B'$, and $C'$ -- is proved in Lemma 4 of \cite{largeKernels-Marco}. 

A simple counting over the binary subspaces of dimension $\ell$ shows that, with high probability, none of the column permutations of $K$ is upper triangular (cf. (47) in \cite{largeKernels-Marco}). Hence, with high probability, $K$ is polarizing \cite{korada2010polar}, which implies that 
\begin{equation}\label{eq:sumABC}
    A'+B'+C'=\lim_{m'\to\infty}\mathbb P\left(Z_{m+m'}\le \ell^{-m'}\right)=1-\epsilon.
\end{equation}
In order to upper bound $B'$, we use the definitions \eqref{eq:ABC} and \eqref{eq:ABCp} of $B$ and $B'$ as well as \eqref{eq:bdmiddle}:
\begin{equation}\label{eq:ubBp}
    B'\le B\le \left(c_3\ell^{-1/2+10/\ln \ell}\ln \ell\right)^m.
\end{equation}
Furthermore, by using the definition \eqref{eq:ABCp}, we can upper bound $C'$ as
\begin{equation}\label{eq:Cp}
    C'\le \lim_{m'\to\infty}\mathbb P\left(Z_{m+m'}\le \ell^{-m'}\mid Z_m\in ( 1-\ell^{-10m}, 1)\right).
\end{equation}
As the kernel $K$ is polarizing (with high probability), the RHS of \eqref{eq:Cp} equals the capacity of a BEC with erasure probability at least $1-\ell^{-10m}$. Therefore,
\begin{equation}\label{eq:ubCp}
C'\le \ell^{-10m}.    
\end{equation}
As a result, we conclude that $A=\mathbb P(Z_m\in [0, \ell^{-10m}))$ is lower bounded as
\begin{equation}\label{eq:lch}
\begin{split}
    A&\ge A'=1-\epsilon-B'-C'\\
    &\ge 1-\epsilon-\left(c_3\ell^{-1/2+10/\ln \ell}\ln \ell\right)^m-\ell^{-10m},
\end{split}
\end{equation}
where the equality in the first line follows from \eqref{eq:sumABC}, and the inequality in the second line follows from \eqref{eq:ubBp} and \eqref{eq:ubCp}. The desired result is readily implied by \eqref{eq:lch}.
\end{IEEEproof}

\end{document}